# DIGITAL IMAGE DATA HIDING TECHNIQUES: A COMPARATIVE STUDY

Minati Mishra[1], Priyadarsini Mishra[2] and Flt. Lt. Dr. M.C. Adhikary[3]

*With the advancements in the field of digital image processing during the last decade, digital image data hiding techniques such as watermarking, Steganography have gained wide popularity. Digital image watermarking techniques hide a small amount of data into a digital image which, later can be retrieved using some specific retrieval algorithms to prove the copyright of a piece of digital information whereas, Steganographic techniques are used to hide a large amount of data secretly into some innocuous looking digital medium. In this paper we are providing an up-to-date review of these data hiding techniques.*

## 1. Introduction

The popular saying 'a picture is worth a thousand words' was certainly true until last decade but, the growing research interests in the field of digital image processing during the last decade have changed this estimation about a picture. Now pictures in their digital representations speak much more than a thousand words, thanks to the digital image data hiding procedures. For example, the block diagram given in figure.1 explains the process of Steganography in which we generally embed some secret message into an innocuous looking simple image (called as the cover image) and create a Stego image. The Stego image visually seems to be indifferent from the original cover but hides the secret message inside it and is transmitted to the desired recipients over the communication channels without creating any suspicion in the minds of the intermediately sniffers or/and receivers. When the authorised recipient receives the image, they follow the extraction procedure to retrieve the secret message. To increase the secrecy or security of the hidden message there may some keys involved in this process of embedding and extraction. At the transmission end, during embedding, the message can suitably be encrypted using one or more encryption techniques. These encryption standards can be key based encryptions or non-key based and in key based techniques, they again can be public or private or a mix. Depending upon the encryption method used during the embedding process, the receiver needs to execute certain decryption algorithms to retrieve the correct message. If any of the decryption algorithms or the keys used for the procedure or the sequence is not known to the receiver then the extraction fails and the receiver cannot retrieve the message.

[1,3]Fakir Mohan University, Balasore, Odisha
[2]DRDA, Balasore, Odisha

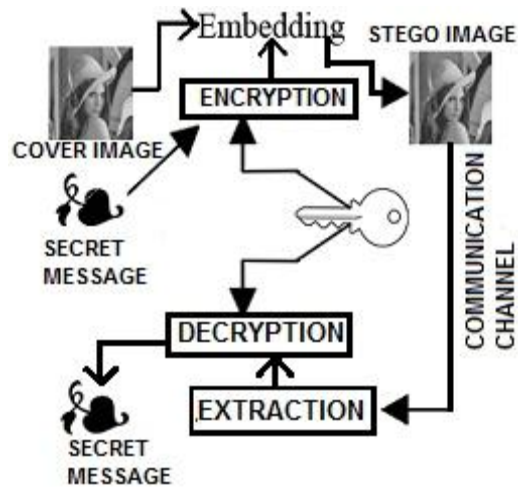

**Fig.1:** Block Diagram of Digital Image Steganography

The above description is all that figure.1 speaks to an ordinary viewer in general. But, does it convey something beyond all these? Does it say something about the authors of this paper? Does it say something about hidden communication methods other than Steganography? These questions can be answered by opening the above figure in notepad. The secret hidden in figure.1 is given in figure.2 which is obtained by opening the above figure in notepad.

```
stegoblock - Notepad
File  Edit  Format  View  Help

============
This Block diagram is created by: MiSS Minati Mishra,Lecturer, P.G. Department of
Information and Communication Technology,Fakir Mohan University,Balasore, ODISHA,
INDIA
to be used as a BLOCK Diagram in her Steganography papers
Description
================
This is my first and simplest demonstration of Steganography: The art of secret
messaging!As u all might have heard of this term Steganography a number of times,
r u curious to know about this? Ok let me define this term for u all..
Definition:
Steganography is an art of hiding information in ways that prevent the detection of
hidden messages and this is achieved by hiding a piece of information (secret message
inside another piece of innocent looking information (Cover).The cover medium can be
text, image, audio or video.
There are a number of covert communication techniques such as:
==>Cryptography, ===>Steganography,==>Covert channel,==>Anonymity,==>Watermarking etc

Steganography is one of the effective means of data hiding that protects data from
unauthorized or unwanted disclosure. It works by hiding secret messages into ordinary
and innocent looking messages those are generally out of suspicion.Out of all other
types of steganograpic procedures, Digital Image Steganography is more popular.

Digital image Steganography procedures exploit the high capacity and widely used
digital images for data hiding purposes.
```

**Fig.2**: figure1 opened in notepad

[1,3]Fakir Mohan University, Balasore, Odisha
[2]DRDA, Balasore, Odisha

According to figure 2, figure 1 is saying many things other than that specified in our discussion above. Does it a magic? No, this is not magic but what Steganography is! This is a simple and small example of digital image Steganography using which one can hide a huge amount of text after the EOF of any JPEG or other image files. The embedded information does not do any visual distortion to the image as the imaging software/ tools do not read beyond EOF but when someone tries to read the picture in notepad, they get the text message hidden inside beyond the EOF of the image files [1]. This technique can be applied to images of all formats and a lot of textual information can be hidden into any given image without doing slightest alteration to the visual quality of the image.

Continuing further with our exploration to discover what else figure.1 can reveal let's bit slice the image. By bit slicing the image, we can discover the following sets of text and images as given in fig.3, in three lower order bit planes of figure 1. This small gray scale image of size 256x256 bits presently holds all these information inside it without giving slightest of hint about all these and without any visible degradation to the picture quality. Interestingly, this is again not all about its capacity and further more information can be embedded making use of other suitable algorithms.

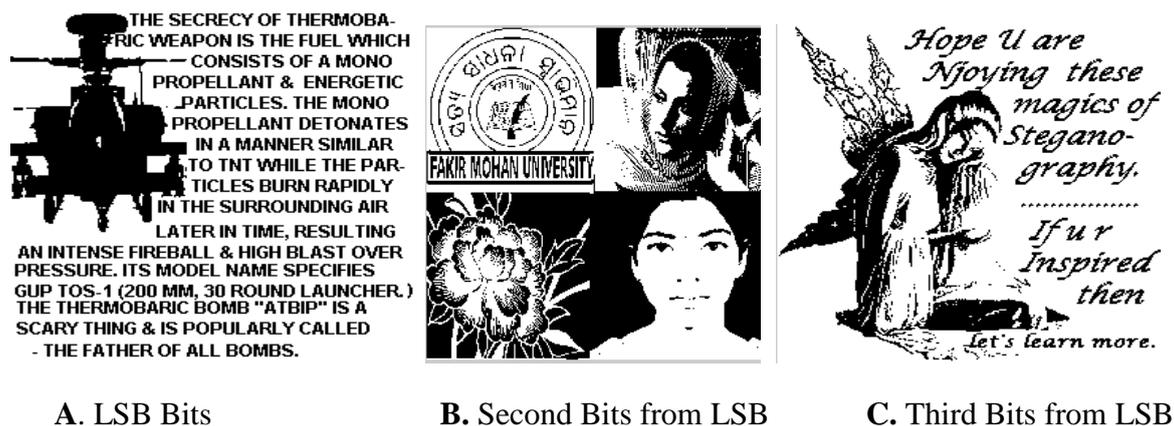

**A**. LSB Bits  **B.** Second Bits from LSB  **C.** Third Bits from LSB

**Fig.3.** Three Least significant Bit planes of Fig.1

The objective of such an elegant and lengthy introduction and this reverse presentation (from results and discussions towards details rather than the usual format of details to experiment and discussion) was to create interest and enthusiasm among the readers about digital image Steganography. With an assumption that our objective is fulfilled and we have inspired at least a few to work in this field, now let us discuss about the types of data hiding techniques, define the term Steganography, discuss about its history in the subsequent sections.

## 2. Types of Data Hiding Techniques

Information hiding techniques are broadly classified into four categories such as, Covert channels, Steganography, Anonymity and Copyright marking [1-5]. The Steganographic procedures can be linguistic or technical whereas the copyright marking procedures can be robust or fragile. Watermarking is a type of robust copyright marking technique which can further be classified as perceptible or imperceptible watermarking. Figure.4 gives a complete classification of various data hiding techniques [5].

[1,3]Fakir Mohan University, Balasore, Odisha
[2]DRDA, Balasore, Odisha

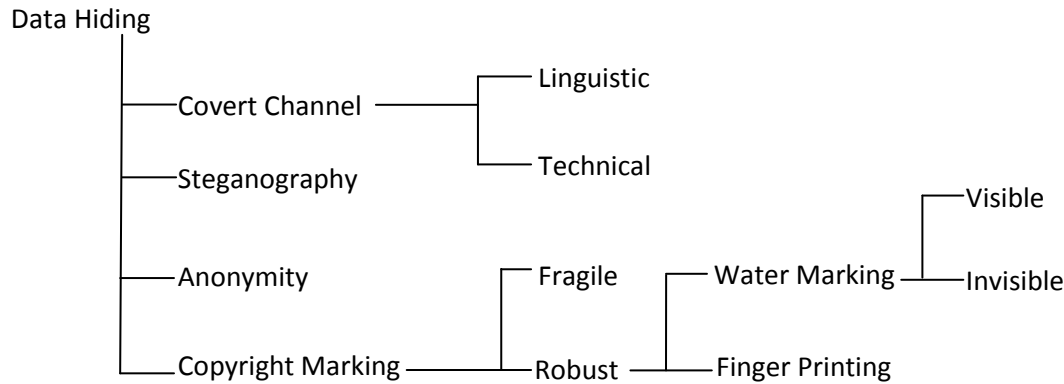

**Fig.4.** Data hiding techniques

A *covert channel* is a type of computer security attack [6] that provides a channel for transfer of information in a way that violates the computer security policy. Robustness and imperceptibility are the important characteristics of a covert channel.

*Linguistic Steganography* (Text Steganography [7-8] or Cryptography) uses text as the cover media to hide the secret message whereas the *technical covert channels* work by exploiting the loopholes in the OS, network model, protocols etc.

*Copyright marking* is a procedure that is used to protect the intellectual properties. In this method a logo or a mark is embedded into a piece of information to show the originality of the work. The copyright can be robust or fragile depending upon the requirement. *Fragile copyright marks* are used to prove manipulations as the fragile marks cannot resist manipulations and lost upon slightest modifications. *Robust copyright methods* are resistant against all sorts of statistical and other types of manipulations. Finger printing and watermarking techniques [9] are popular types of robust copyright marking methods and are used for authentication purposes [10]. Table 2 compares watermarking against Steganography. Figure.6 shows the experimental results of spatial domain visible watermarking in which a 32X32 pixel monochrome watermark is embedded into the higher order bit plane of the 256X256 pixel gray scale Woman image and then the watermark is retrieved after the watermarked image is subjected to various types of statistical attacks such as format change, resizing, compression, rotation etc. In Figure.7, a 20x20 pixel gray scale watermark is embedded into the woman image invisibly into random locations using the Fibonacci-Lucas transformation [11] and then the watermark is successfully retrieved after the image is subjected to different attacks. The Fibo-Lucas transformation, in this experiment, ensures security of the watermark against unauthorised retrieval/modification along with the other desirable properties. The results of the experiments show that the procedures are robust against all the attacks.

*Anonymity* is a method of secret communication where the transmitter and the receiver remain anonymous so that a third party, who is interested on the information but is not a legitimate user of the information, looses track of it.

[1,3]Fakir Mohan University, Balasore, Odisha
[2]DRDA, Balasore, Odisha

# 3. Steganography

Steganography is a process of secret communication where a piece of information (a secret message) is hidden into another piece of innocent looking information, popularly called a cover, in such a way that the very existence of the secret information remains concealed without raising any suspicion in the minds of the viewers.

## Description of various terms

*Cover*: The information/ Image that camouflages the secret message inside.

*Embedding and Extraction Processes* are the processes used to hide the secret message inside the Cover and retrieve the hidden information from the Stego medium respectively.

*Encryption*: This embedding process can suitably be preceded by a phase of encryption where the secret message is coded to a piece of intangible medium.

*Stego information*: The resultant information after embedding is known as Stego information.

*Decryption Process*: The extraction phase can be followed by a suitable decryption phase depending upon whether encryption was used or not during embedding. In this phase the encoded piece of intangible information is converted back to the tangible format with the use of a suitable decoding algorithm.

Steganography can be classified into several sub classes such as *text, audio, video* or *image* Steganography, depending upon whether text, audio, video or image is used as the cover medium. Out of all these available forms of Steganography, digital image Steganographic procedures are more popular among the researchers as images are more common forms of mediums that are used worldwide for data transmission and also due to their data hiding capacity. Before discussing further about this form of Steganography, let's first define what a digital image is.

**Definition 1:** A digital image is a two dimensional function $f(x, y)$ where, x and y are spatial coordinates, $f$ is the amplitude at (x, y), also called the intensity or gray level of the image at that point and x, y, $f$ are finite- discrete quantities[12].

These digital images can be monochrome (bi-tone), grayscale or color depending upon the permissible intensity levels of each pixel i.e. whether each pixel is represented by only one bit, 8-bits or 24-bits. Generally speaking, a monochrome image can have only one bit plane whereas there are 8-bit planes in a gray scale image and 24 bit planes (8-bits each, with respect to the 3 channels R, G and B) in a color image. Figure 5 shows the bit planes of a 7X8 pixel gray scale image. The least significant bit plane (LSB plane) is the plane that consists of bits with minimum positional value ($2^0$= 1) and the MSB plane (most significant bit plane) consists of bits with highest positional values i.e. $2^7$=128.

**Definition 2:** Digital Image processing is the process that uses computer algorithms to perform image processing on digital images. It allows a wider range of complex and sophisticated algorithms to be applied to digital images with ease and with a much effective way in comparison to analog signal processing.

[1,3]Fakir Mohan University, Balasore, Odisha
[2]DRDA, Balasore, Odisha

**Definition 3:** The embedding and extraction processes in digital image Steganography are mapping given by:

$$M : \{C \times (K) \times T\} \rightarrow C' \quad \text{---------- (eq. 1)}$$

And

$$X : \{C' \times (K)\} \rightarrow C \quad \text{--------- (eq. 2)}$$

Where, M and X are the embedding and extraction functions respectively, C is the cover image, C' is the stego image, K is an optional set of keys (K in M and N may be same or may be different depending upon the encryption algorithm used), T is a set of secret messages.

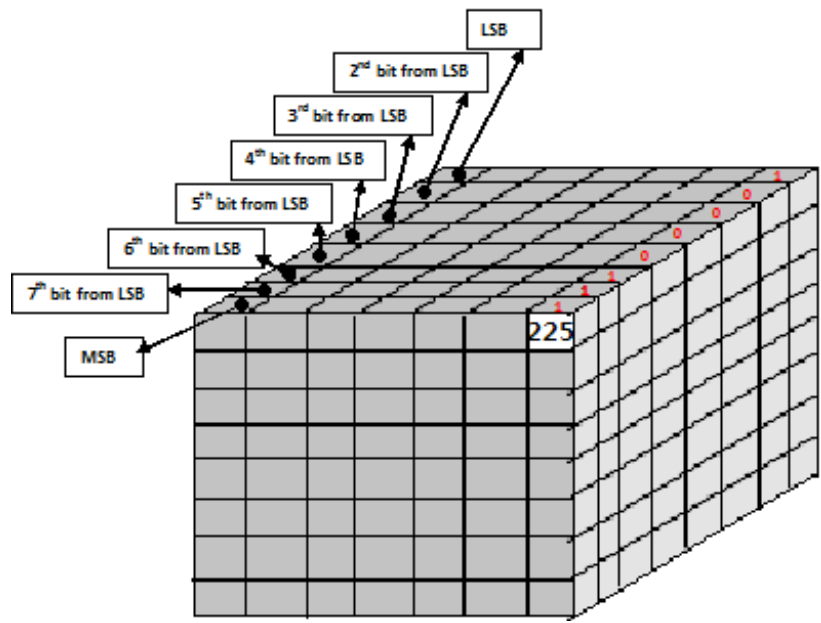

**Fig.5.** Bit planes of a 7X8 pixel gray scale image

## 3.1 Attributes of Steganography

The desired attributes of Steganography are imperceptibility, capacity and robustness [1].

A. **Imperceptibility**: This is the first and foremost requirement of any Steganographic algorithm. According to this feature, a good Steganographic system should not cause any degradation to the quality of the stego, the secret message should remain invisible and undetectable to normal human vision and there should be no visual difference between the original object and the stego object so as to make it unsusceptible and safe.

B. **Capacity** – The information hiding capacity of Steganography should be very high in contrast to that of watermarking, which needs to embed only a small amount of copyright information.

---

[1,3]Fakir Mohan University, Balasore, Odisha
[2]DRDA, Balasore, Odisha

C. **Robustness** – the system should be robust against statistical attacks and/or image manipulations. Statistical steganalysis is the practice of detecting hidden information through applying statistical tests on image data. Many Steganographic algorithms leave a "signature" when embedding information that can be easily detected through statistical analysis. To be able to pass by a warden without being detected, a Steganographic algorithm must not leave such a mark in the image as be statistically significant.

Unfortunately, as on date, there exists no Steganography system that satisfies all the three requirements and research is still going on in this direction.

## 3.2 Types of Steganography

Steganography Techniques are broadly classified into two categories such as spatial domain techniques and transform domain techniques. The more popular spatial domain methods take advantage of the human visual system and directly embed data by manipulating the pixel intensities. In transform domain procedures, the image is first transformed into frequency domain and then the message is embedded. The transform domain procedures are more robust against statistical attacks and manipulations in comparison to the spatial domain methods but spatial domain techniques are more popular due to their simplicity and ease of use.

Depending upon the embedding and extraction procedures used Steganographic systems can again be classified into the following three different categories [13]:

A. **Pure Steganography (or No Key Steganography - NKS)**: This is the simplest and weakest form of Steganography in which the secret message is directly embedded into the cover image without any encryption. The success of this hidden communication depends upon the assumption that parties other than the intended receivers (attackers) are not aware of the existence of the secret message within.

B. **Secret Key Steganography (SKS)**: In this form of Steganography, both the receiver and transmitter have common agreed upon secret keys. The secret message is embedded into and extracted out of the stego image using these keys. The keys can be separately shared between both parties using some confidential channel prior to the actual transmission starts. The strength of this system is its higher security. Parties other than the intended receiver cannot retrieve the secret message or will require very high computational time and power to retrieve it applying some brute force methods, in case they suspect the presence of the secret information. The robustness of this system, of course, lies with the secrecy of the keys and the difficult part in this method is how to share the keys between the transmitting and receiving parties maintaining their secrecies.

C. **Public Key Steganography (PKS)**: This methods use a pair of public and private keys to hide the secret information. The key benefits of this system are its robustness as well as easy key management. The method is robust because the parties other than the intended receivers need to know both the private and public keys used for embedding and the encryption algorithms used, in order to be able to extract the hidden information.

[1,3]Fakir Mohan University, Balasore, Odisha
[2]DRDA, Balasore, Odisha

## 4. History of Steganography

The history of Steganography can be guessed to be as old as history of writing. It has been in use in some form or other form approximately since last 3000 years. The word Steganography has come from two Greek words: 'Stegein' means, 'to cover' and 'Grafein' means, "to write". Hence Steganography literally means 'covered writing'. It is an effective means of hiding data, thereby protecting the data from unauthorized or unwanted viewing. It hides secret messages in innocent message, which looks like ordinary message and would not raise suspicion. An examples of Steganography is null ciphers, in this the secret messages are hidden inside a normal message, which looks like an ordinary message. It has been used in various forms for thousands of years. In the 5th century BC, Histaiacus shaved a slave's head, tattooed a message on his skull and the slave was dispatched with the message after his hair grew back. In Saudi Arabia at the King Abdulaziz City of science and technology, a project was initiated to translate into English some ancient Arabic manuscripts on secret writing which are believed to have been written1200 years ago. Some of these manuscripts were found in Turkey and Germany. Five hundred years ago, the Italian mathematician Jérôme Cardan invented a Chinese ancient method of secret writing. The scenario goes as follows: a paper mask with holes is shared among two parties, this mask is placed over a blank paper and the sender writes his secret message through the holes then takes the mask off and fills the blanks so that the message appears as an innocuous. This method is credited to Cardan and is called Cardan Grill. It was also reported that the Nazis invented several Steganographic methods during World War II such as Microdots, and have reused invisible ink and null ciphers [14].

However today, Steganography has gone significantly more sophisticated than the example given above allowing a user to hide large amounts of information into image, text, audio and video files.

## 5. Comparison between different Data Hiding Techniques

In our first example of hiding textual information into an image after the EOF of the image, we have followed the simple copy command of DOS. In this method we have tried hiding as many as 100 text files of total size 20 Mb in a small bmp file of 2.06 kb without making any visual distortion to the original image. The advantages of this technique are its simplicity, high imperceptibility and high capacity. Data can be hidden inside all types of image formats using this method but the disadvantage is; this is a very fragile technique and does not tolerate any manipulation [6]. Even slightest modification to the image or change of format destroys the hidden data.

In the second example, a little robust method is used for data hiding. It is known that every pixel in a gray scale image can have a value between 0 and 255 and when all the three least significant bits of a pixel are altered, the pixel value is changed maximum by +7 to – 7 (refer table 1) and it has been seen that these substitution methods do not alter all the bits of the original image. It is found that only a maximum of 50-55% bits get affected though this method and rest of the bits remain unchanged preserving the quality of the image. Since the human vision (HVS) cannot detect this small change in intensity therefore, it is being exploited for the purpose of data hiding. The image is bit sliced and the lower order bits, those do not carry much important picture information, are replaced by suitable message bits and then again the image is reconstructed. It has been seen that the Peak Signal to Noise Ration (PSNR) value remains more than 30dB which is within the acceptable range, even after 4 image bits are replaced by message bits using this method [15-16]. A comparison between various data hiding techniques is provided in table 3.

[1,3]Fakir Mohan University, Balasore, Odisha
[2]DRDA, Balasore, Odisha

**TABLE 1**: Difference in pixel intensity with 3LSB substitution

| Actual Bit pattern | Altering 3 LSBs | Difference in intensity |
|---|---|---|
| xxxxx000 | xxxxx111 | -7 |
| xxxxx001 | xxxxx110 | -5 |
| xxxxx010 | xxxxx101 | -3 |
| xxxxx011 | xxxxx100 | -1 |
| xxxxx100 | xxxxx011 | 1 |
| xxxxx101 | xxxxx010 | 3 |
| xxxxx110 | xxxxx001 | 5 |
| xxxxx111 | xxxxx000 | 7 |

*x can be either zero or one. The 3LSB substitution method does not do any modification to the higher order bits.

**Fig.6. Visible watermarking and watermark retrieval after different statistical attacks**

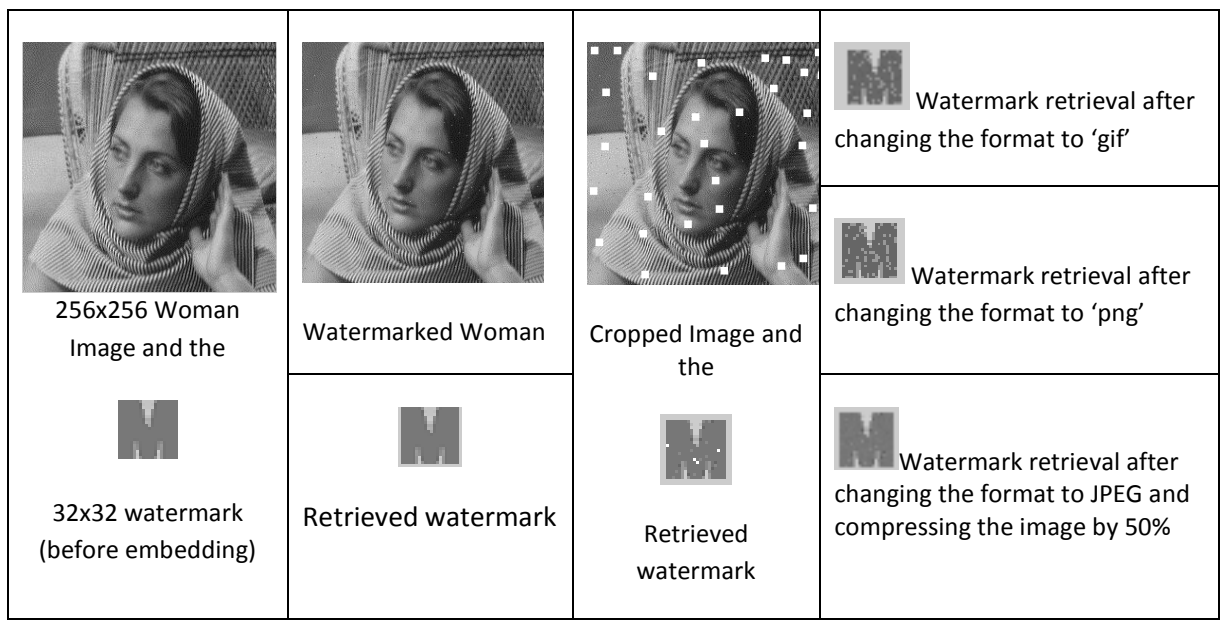

**Fig.7. Invisible Watermarking and watermark retrieval after different attacks**

**TABLE 2**: Watermarking versus Steganography

| Technique | Attributes/ Expected values | Use of the technique | Types | Characteristics |
|---|---|---|---|---|
| Steganography | • Capacity<br>• Robustness<br>• Message Secrecy<br>• Impeccability - Should be high for all these attributes | For confidential data transmission such as Defence Data, Medical history of patients. | Special domain | • Procedural Simplicity & ease of use.<br>• Capacity & Impeccability are high.<br>• Message Secrecy is very high in key based methods but robustness against statistical attacks is generally low. |
|  |  |  | Transform domain | • Difficult to implement.<br>• Capacity is low in comparison to Spatial Domain.<br>• Message Secrecy, Impeccability and Robustness are high. |

[1,3] Fakir Mohan University, Balasore, Odisha
[2] DRDA, Balasore, Odisha

| Watermarking | • Robustness<br>• Message Secrecy<br>   - Should be high<br>• Capacity- not important as a small amount of data need to be embedded<br>• Impeccability – Depends | For Copyright authentication | Visible | • Embedding is generally performed into the higher order bit planes.<br>•Impeccability is low as the watermark is visible.<br>•Robustness is high |
|---|---|---|---|---|
| | | | Invisible | •Embedding is generally performed into selected pixels of the whole image.<br>•Impeccability is high.<br>•Robustness is high.<br>•High security against unauthorised alteration of the watermark. |

**TABLE 3**: Comparison between different data hiding Procedures

| Technique | | Hiding Capacity | Robustness | Message Secrecy | Imperceptibility | Key management/ Embedding Complexity |
|---|---|---|---|---|---|---|
| After EOF | | Very High | Very low | Low | Very High | Not required/ Very Low |
| NKS | OBS | Low | medium | Low | Very High | Not required/ Low |
| | MBS | High | | | Very High upto 3 bits substitution | Not required /Low |
| PKS | OBS | Low | medium | High | Very High | Easy/ Low |
| | MBS | High | | | Very High upto 3 bits substitution | Easy / Low |
| SKS | OBS | Low | medium | Very High (Lies with the secrecy of the Keys) | Very High | Difficult /Low |
| | MBS | High | | | Very High upto 3 bits substitution | Difficult / Low |
| TDS | | Medium | High | Very High | Very High | High |
| VWM | | Low | High | N/A | N/A | High |
| IVWM | | Low | High | High | Very High | High |

*OBS: One Bit Substitution
MBS: Multiple Bit Substitution
TDS: Transform Domain Steganography
VWM: Visible Watermarking
IVWM: Invisible Watermarking

# 6. Conclusion

Like every coin has two sides, all technological developments are associated with both bad as well as good applications and Steganography is not an exception to this. Though there are many good reasons to use data hiding techniques and these should be used for legitimate applications only but, unfortunately, Steganography can also be used for illegitimate reasons. For instance, someone trying to steal data can conceal it in another file and send it out in an innocent looking email. The information stolen and passed may be a patient's confidential test reports, the tender information of a company/ organization or even the defence plans of a country. No doubt, terrorists and criminals can use this method to secretly spread their action plans and though no evidence is yet established to this claim still, it is claimed that, Steganography was used to pass the execution plan of the 9/11 WTC attack. Therefore, through this paper we have just tried to create awareness and establish the fact that such methods do exist. Still researches are on progress all over the world on this innovative field of Steganography, keeping in mind the positive impacts of this on the society in the present ICT based revolutionary age of 21st century.

[1,3]Fakir Mohan University, Balasore, Odisha
[2]DRDA, Balasore, Odisha

## References:

1. A. Cheddad et al.: "Digital image Steganography: Survey and analysis of current methods", Signal Processing, Elsevier, 90(2010) 727-752.

2. B. Pfitzmann: "Information hiding terminology", Information Hiding: First International Workshop, Cambridge, U.K., May 30 - June 1, 1996. Proceedings (Lecture Notes in Computer Science) , pp. 347-350, ISBN 3-340-61996-8.

3. T. Aura, "Invisible Communication", EET 1995, technical report, Helsinki Uni. Of Technology, Finland.

4. W. Bender et al., "Techniques for Data Hiding", IBM Systems, Vol.35, No.3 & 4, 1996, pp. 313-336.

5. Fabien A.P. Petitcolas et al., "Information Hiding – A survey", Proceedings of IEEE, Special Issue on protection of multimedia content, 87(7):1062-1078, July 1999.

6. http://en.wikipedia.org/wiki/Covert_channel

7. M.H. Shirali-Shahrez, M. Shirali-Shahrez, "A New Synonym: Text Steganography", Intl Conf. on Intl Info. Hiding and Multimedia Signal Processing, 978-0-3278-3/08, 2008 IEEE.

8. Kan Farhan Rafat et al: "Survey Report- State of the art in digital Steganography focussing ASCII text Documents", IJCSIS, Vol.7, No.2, 2010.

9. Vidyasagar M. Potdar et al., "A Survey of Digital watermarking Techniques", 3$^{rd}$ IEEE Intl Conf. on Industrial Information (INDIN), pp. 709- 716, 2005 IEEE.

10. M.D. Swanson et al., "Robust Data hiding for images", 7$^{th}$ Digital Signal Processing Workshop (DSP 96), pp. 37-40, IEEE, Loen, Norway, Sep. 1996.

11. Minati Mishra et al., "IMAGE ENCRYPTION USING FIBONACCI-LUCAS TRANSFORMATION", International Journal on Cryptography and Information Security (IJCIS), pp. 131-141, Vol.2, No.3, September 2012.

12. RC Gonzalez, RE Wood: Digital Image Processing, 2nd Ed, PHI,    New Delhi, 2006.

13. Neil F. Johnson: "Exploring Steganography: Seeing the Unseen", George Mason University, IEEE Computer, pp. 26-34, Feb 1998.

14. Bin Li et al., "A Survey on Image Steganography and Steganalysis", Journal of Info. Hiding and Multimedia Signal Processing, ISSN 2073-4212, Vol-2, No-2, pp142-172, Apr 21011.

15. Minati Mishra, A.R. Routray, M.C. Adhikary, "Secured Steganography with Image Encryption through Chaotic Mapping", ANVESA, Vol-6, Issue 1&2, pp. 7-11, December 2011.

16. Minati Mishra, Sunit Kumar and Subhadra Mishra: "Security Enhanced Digital Image Steganography Based on Successive Arnold Transformation", Advances in Intelligent and Soft Computing, Springer 2012, Volume 167/2012, pp. 221-229, DOI: 10.1007/978-3-642-30111-7_21.

[1,3]Fakir Mohan University, Balasore, Odisha
[2]DRDA, Balasore, Odisha